  \documentstyle[aas2pp4,epsf]{article}

\newcommand{\Ell}{E_\parallel}      
\newcommand{\Bc}{B_{\rm cnt}}      
\newcommand{\sgT}{\sigma_{\rm T}}  
\newcommand{\sgP}{\sigma_{\rm p}}  
\newcommand{\rlc}{\varpi_{\rm LC}} 
\newcommand{\Rc}{R_{\rm C}}        
\newcommand{\Ex}{\epsilon_{\rm x}} 
\newcommand{\Eg}{\epsilon_\gamma}  
\newcommand{\inc}{\alpha_{\rm i}}  

\lefthead{Hirotani and Shibata}
\righthead{}

\begin{document}

\title{Electrodynamic Structure of an Outer--Gap Accelerator:
       Location of the Gap and the Gamma-Ray Emission from the Crab Pulsar}
\author{Kouich Hirotani}
\affil{National Astronomical Observatory, 
       Mitaka, Tokyo 181-8588, Japan\\
       hirotani@hotaka.mtk.nao.ac.jp}
\and
\author{Shinpei Shibata}
\affil{Department of Physics, Yamagata University,
       Yamagata 990-8560, Japan\\
       shibata@sci.kj.yamagata-u.ac.jp}

\begin{abstract}
We investigate a stationary pair production cascade 
in the outer magnetosphere of a spinning neutron star.
The charge depletion due to global flows of charged particles,
causes a large electric field along the magnetic field lines.
Migratory electrons and/or positrons are accelerated by this field
to radiate curvature gamma-rays, 
some of which collide with the X-rays to materialize as pairs in the gap.
The replenished charges partially screen the electric field, 
which is self-consistently solved together with the distribution functions
of particles and gamma-rays.
If no current is injected at either of the boundaries of the accelerator,
the gap is located around the conventional null surface,
where the local Goldreich-Julian charge density vanishes.
However, we first find that the gap position shifts outwards (or inwards)
when particles are injected at the inner (or outer) boundary.
Applying the theory to the Crab pulsar, 
we demonstrate that 
the pulsed TeV flux does not exceed the observational upper limit 
for moderate infrared photon density 
and that the gap should be located near to or outside of the 
conventional null surface
so that the observed spectrum of pulsed GeV fluxes may be 
emitted via a curvature process.
Some implications of the existence of a solution for 
a super Goldreich-Julian current are discussed.
\end{abstract}



\section{Introduction}
\label{sec:intro}

The EGRET experiment on the Compton Gamma Ray Observatory
has detected pulsed signals from seven rotation-powered pulsars
(e.g., for Crab, Nolan et al. 1993, Fierro et al. 1998).
The modulation of the $\gamma$-ray light curves at GeV energies 
testifies to the production of $\gamma$-ray radiation in the pulsar 
magnetospheres either at the polar cap 
(Harding, Tademaru, \& Esposito 1978; Daugherty \& Harding 1982, 1996;
 Sturner, Dermer, \& Michel 1995;
 Shibata, Miyazaki, \& Takahara 1998),
or at the vacuum gaps in the outer magnetosphere
(Cheng, Ho, \& Ruderman 1986a,b, hereafter CHR;
 Chiang \& Romani 1992, 1994; Romani and Yadigaroglu 1995;
 Romani 1996; Zhang \& Cheng 1997, ZC97).
Effective $\gamma$-ray production in a pulsar magnetosphere
may be extended to the very high energy (VHE) region above 
100 GeV as well;
however, the predictions of fluxes by the current models of 
$\gamma$-ray pulsars are not sufficiently conclusive
(e.g., Cheng 1994).
Whether or not the spectra of $\gamma$-ray pulsars continue up to the
VHE region is a question that remains one of the 
interesting issues of high-energy astrophysics.

In the VHE region,
positive detections of radiation at a high confidence 
level have been reported from the direction of the Crab pulsar
(Nel et al. 1993).
However, as for {\it pulsed} TeV radiation,
only the upper limits have been, as a rule, obtained
(Akerlof et al. 1993; Borione et al. 1997; Srinivasan et al. 1997;
 Yoshikoshi et al. 1997; Sako et al. 2000).
If the VHE emission originates in the pulsar magnetosphere,
a significant fraction of it can be expected to show pulsation.
Therefore, the lack of {\it pulsed} TeV emissions provides a
severe constraint on the modeling of particle acceleration zones
in a pulsar magnetosphere.

In fact, in the CHR picture,
the magnetosphere should be optically thick for pair--production
in order to reduce the TeV flux to an unobserved level 
by absorption.
This in turn requires very high luminosities of infrared photons.
However, the required IR fluxes are generally orders of magnitude
larger than the observed values (Usov 1994).
We are therefore motivated by the need to contrive an outer--gap model
that produces less TeV emission with a moderate infrared luminosity.

High-energy emission from a pulsar magnetosphere,
in fact, crucially depends on the acceleration electric field, 
$\Ell$, along the magnetic field lines.
It was Hirotani and Shibata (1999a,b,c; hereafter Papers~I, II, III),
and Hirotani (2000a; hereafter Paper~VI)
who first considered the spatial distribution of $\Ell$ 
together with particle and $\gamma$-ray distribution functions.
By solving these Vlasov equations,
they demonstrated that 
a stationary gap is formed around the conventional null surface 
at which the local Goldreich--Julian charge density, 
\begin{equation}
  \rho_{\rm GJ}= -\frac{\Omega B_z}{2\pi c},
  \label{eq:def_rhoGJ}
\end{equation}
vanishes,
where $B_z$ is the component of the magnetic field along 
the rotation axis,
$\Omega$ the angular frequency of the neutron star,
and $c$ the speed of light.
Equation (\ref{eq:def_rhoGJ}) is valid unless the gap is 
located close to the light cylinder,
of which distance from the rotation axis is
given by $ \rlc= c / \Omega$.
The electrodynamic model developed in this paper is basically
the same as Paper~VI.
However, we find an interesting behavior of the gap position,
by relaxing the boundary conditions to allow electric current 
injection through the inner or the outer boundaries of the gap.

Subsequently, Hirotani (2000b, hereafter Paper~IV; 2001, Paper~V) 
considered the \lq gap closure condition' so that
a gap may maintain a stationary pair-production cascade.
In this paper, 
this closure condition is generalized into the case
when the currents are injected through the boundaries.

In the next two sections, we describe the physical processes of pair 
production cascade and the resultant $\gamma$-ray emission.
We then apply the theory to the Crab pulsar 
and present the expected $\gamma$-ray spectra in \S~\ref{sec:app_Crab}.
In the final section, we discuss the possibility of a gap formation
for a super Goldreich-Julian current. 

\section{Analytic Examination of the Gap Position}
\label{sec:analytic}

Let us first consider the gap position analytically
when there is a current injection into the gap.
We consider the particle continuity equations 
in \S~\ref{sec:cont}
and the $\gamma$-ray Boltzmann equations
in \S~\ref{sec:Boltz_gamma}.

\subsection{Particle Continuity Equations}
\label{sec:cont}

Under the mono-energetic approximation, 
we simply assume that the electrostatic and the 
curvature-radiation-reaction forces cancel each other
in the Boltzmann equations of particles.
Then the spatial number density of the outwardly and inwardly
propagating particles, $N_+(s)$ and $N_-(s)$, 
at distance $s$ from the neutron-star surface along the last-open 
field line,
obey the following continuity equations:
\begin{equation}
  \frac{\partial N_\pm}{\partial t} 
  + \vec{v} \cdot \frac{\partial N_\pm}{\partial \vec{x}}
  = Q(\vec{x}),
  \label{eq:cont-eq-0}
\end{equation}
where
\begin{equation}
  Q(\vec{x}) \equiv
  \frac{1}{c} \int_{0}^\infty d\epsilon_\gamma \, 
    [ \eta_{\rm p+} G_+   +\eta_{\rm p-} G_- ];
\end{equation}
$G_+(\vec{x},\epsilon_\gamma)$ and $G_-(\vec{x},\epsilon_\gamma)$ 
refer to the distribution functions of
outwardly and inwardly propagating $\gamma$-ray photons,
respectively, having energy $m_{\rm e}c^2 \epsilon_\gamma$.
The pair production rate for an outwardly propagating
(or inwardly propagating) $\gamma$-ray photon to materialize 
as a pair per unit time is expressed by 
$\eta_{{\rm p}+}$ (or $\eta_{{\rm p}-}$).
For charge definiteness, we consider that 
a positive electric field arises in the gap.
In this case, 
$N_+$ (or $N_-$) represents the number densities of
positrons (or electrons).

The particle velocity at position ($r$,$\theta$) becomes
(eq.~[21] in Paper~VI)
\begin{equation}
  \vec{v}=  \vec{v}_{\rm p} 
          + (r\Omega\sin\theta +\kappa B_\phi -c\Ell \frac{B_\phi}{B^2})
            \vec{e}_\phi,
  \label{eq:velocity}
\end{equation}
where $\kappa$ is a constant and $\vec{e}_\phi$ refer to
the azimuthal unit vector.
In the parentheses,
the term $r\Omega\sin\theta$ is due to corotation,
while $\kappa B_\phi$ due to magnetic bending.
Since $\Ell$ arises in the gap, the corresponding drift velocity
appears as $-c\Ell B_\phi/B^2$.
Unless the gap is located close to the light cylinder, 
we can neglect the terms containing $B_\phi$ as a first--order 
approximation.
We thus have 
\begin{equation}
  \vec{v} \approx \vec{v}_{\rm p} + r\Omega\sin\theta \vec{e}_\phi.
\end{equation}
Imposing a stationarity condition
\begin{equation}
  [\partial_t +(r\Omega\sin\theta)\partial_\phi] N_\pm = 0,  
\end{equation}
reminding that the projected velocity on the poloidal plane is
$\vec{v}_{\rm p} = c\cos\Phi \vec{B}_{\rm p}/B_{\rm p}$,
and utilizing $\mbox{div}\vec{B}_{\rm p} \approx \mbox{div}\vec{B}= 0$, 
we obtain
\begin{equation}
  \pm B \frac{\partial}{\partial s}\left( \frac{N_\pm}{B} \right)
  = \frac{1}{c \cos\Phi} \int_{0}^\infty d\epsilon_\gamma \, 
    [ \eta_{\rm p+} G_+   +\eta_{\rm p-} G_- ],
  \label{eq:cont-eq}
\end{equation}
where $\Phi$ refers to the projection angle of the 
particle three-dimensional motion onto the poloidal plane.
It is defined by
$\Phi = \arcsin(r_{\rm cnt}\Omega\sin\theta / c$), 
where $r_{\rm cnt}$ is the distance of the gap center
from the star center.
The pair production rate per unit time by a single $\gamma$-ray photon,
$\eta_{{\rm p}\pm}$, are defined as
\begin{equation}
  \eta_{{\rm p}\pm}(\Eg)
  = (1-\mu_{\rm c}) c
     \int_{\epsilon_{\rm th}}^\infty d\epsilon_{\rm x}
     \frac{dN_{\rm x}}{d\Ex} 
     \sgP(\Eg,\Ex,\mu_{\rm c}),
  \label{eq:def_etap_0}
\end{equation}
where $\sgP$ is the pair-production cross section and
$\cos^{-1}\mu_{\rm c}$ refers to the collision angle between 
the $\gamma$-rays and the X-rays 
(see Paper~VI for more details about eq.~[\ref{eq:def_etap_0}]);
$\epsilon_{\rm th} \equiv 2/[(1-\mu_{\rm c})\Eg]$.
The adopted value of $\mu_{\rm c}$ will be detailed in \S~\ref{sec:X}.
The quantity $\Ex$ refers to the X-ray energy in the unit of
$m_{\rm e}c^2$. 

Although $\Phi \ne 0$ is adopted after \S~\ref{sec:basic},
in this section we simply neglect the projection effect of the poloidal 
velocity and put $\Phi=0$.
Then equation~(\ref{eq:cont-eq}) gives
\begin{equation}
  \pm B \frac{d}{ds} \left(\frac{N_\pm}{B}\right)
     = \frac{1}{\lambda_{\rm p}}
          \int_0^\infty d\Eg (G_+ +G_-),
  \label{eq:cont-eq2}
\end{equation}
where $G_+(s,\Eg)$ and $G_-(s,\Eg)$ refer to the
distribution functions of the outwardly and inwardly 
propagating $\gamma$-rays;
the mean free path $\lambda_{\rm p}$ is defined by
\begin{equation}
  \lambda_{\rm p} 
  \equiv 
  \frac{1}{c} \frac{\displaystyle \int_0^\infty \eta_{{\rm p}+} G_+ d\Eg}
                   {\displaystyle \int_0^\infty G_+ d\Eg}.
  \label{eq:def_mfp}
\end{equation}
Since $W \ll \rlc$ is justified for the Crab pulsar (Paper~V),  
we regard $\lambda_{\rm p}$ to be constant in the gap in this section.

\subsection{Boltzmann Equations for Gamma-rays}
\label{sec:Boltz_gamma}

Unlike the charged particles,
$\gamma$-rays do not propagate along the magnetic field line at each point,
because they preserve the directional information where they were emitted.
However, to avoid complications, we simply assume 
that the outwardly (or inwardly) propagating $\gamma$-rays
dilate (or constrict) at the same rate with the magnetic field.
This assumption gives a good estimate 
when $W \ll \rlc$ holds.
We then obtain (Paper~VI)
\begin{eqnarray}   
  \pm B \frac{\partial}{\partial s} \left( \frac{G_\pm}{B}\right)
     = - \frac{\eta_{{\rm p}\pm}}{c\cos\Phi} G_\pm
       + \frac{\eta_{\rm c}     }{c\cos\Phi} N_\pm,
  \label{eq:Boltz_gam}
\end{eqnarray}   
where (e.g., Rybicki, Lightman 1979)
\begin{equation}
  \eta_{\rm c} \equiv \frac{\sqrt{3}e^2 \Gamma}{h \Rc}
               \frac1{\epsilon_\gamma} 
	       F \left( \frac{\epsilon_\gamma}{\epsilon_{\rm c}} \right) ,
  \label{eq:def-etaC}
\end{equation}
\begin{equation}
  \epsilon_{\rm c} \equiv \frac1{m_{\rm e}c^2} \frac3{4\pi}
                          \frac{hc \Gamma^3}{\Rc} ,
  \label{eq:def_Ec}
\end{equation}
\begin{equation}
  F(s) \equiv s \int_x^\infty K_{\frac53} (t) dt ;
\end{equation}
$\Rc$ is the curvature radius of the magnetic field lines
and $K_{5/3}$ is the modified Bessel function of $5/3$ order.
The effect of the broad spectrum of curvature $\gamma$-rays
is represented by the factor $F(\epsilon_\gamma/\epsilon_{\rm c})$
in equation (\ref{eq:def-etaC}).

Noting that the absorption due to pair production is negligible 
compared with curvature emission term on the right-hand side
of equation~(\ref{eq:Boltz_gam}),
and putting $\Phi=0$ again, we obtain
\begin{equation}
  \pm \frac{\partial}{\partial s} 
  \left[ \frac{1}{B} G_\pm(s,\Eg) \right] 
  = \frac{\eta_{\rm c}(\Eg)}{c} 
    \frac{N_\pm(s)}{B}.
  \label{eq:Boltz-gam2}
\end{equation}
Integrating equation~(\ref{eq:Boltz-gam2}) over $\Eg$,
and combining with equation~(\ref{eq:cont-eq2}), we obtain
\begin{equation}
  \pm \frac{d^2}{ds^2} \left( \frac{N_\pm}{B} \right)
  = \frac{1}{\lambda_{\rm p} c} \frac{N_+ -N_-}{B}
    \int_{\beta_0}^{\beta_N} \eta_{\rm c}(\Eg) d\Eg,
  \label{eq:cont-eq3}
\end{equation}
where $\beta_N$ is the upper cutoff dimensionless $\gamma$-ray energy.
In the present paper, we set $\beta_N= \beta_9= 10^{5.5}$
(see \S~\ref{sec:basic}).

One combination of the two independent equations~(\ref{eq:cont-eq3})
yields the conserved current per magnetic flux tube,
\begin{equation} 
  \frac{\Omega}{2\pi} j_{\rm tot}
  = ce \frac{N_+(s) +N_-(s)}{B(s)}
  \label{eq:consv_0}
\end{equation}
If $j_{\rm tot}=1.0$, 
the conserved current density becomes its Goldreich-Julian value.
Another combination of equations~(\ref{eq:cont-eq3}) gives
\begin{equation} 
  \frac{d^2}{ds^2} \left(\frac{N_+ -N_-}{B}\right)
  = \frac{4 N_\gamma}{\lambda_{\rm p}} \frac{N_+ -N_-}{B},
  \label{eq:master_eq_1}
\end{equation}
where 
\begin{equation}
  N_\gamma 
  \equiv 
  \frac{W/2}{c} \int_{\beta_0}^{\beta_N} \eta_{\rm c}(\Eg) d\Eg
  \label{eq:Ngamma}
\end{equation}
referes to the expectation value of the number of $\gamma$-rays
emitted by a single particle that runs a typical length $W/2$
in the gap.

In a stationary gap, the pair production optical depth,
$W/\lambda_{\rm p}$, must equal the 
expectation value for a $\gamma$-ray to materialize with the gap,
$N_\gamma^{-1} (j_{\rm gap}$/$j_{\rm tot})$.
We thus obtain the following condition:
\begin{equation}
  W = \frac{\lambda_{\rm p}}{N_\gamma}
      \frac{j_{\rm gap}}{j_{\rm tot}},
  \label{eq:closure}
\end{equation}
which is automatically satisfied by the stationary Vlasov equations. 
Here, the dimensionless current density, $j_{\rm gap}$, 
created in the gap is defined by
\begin{eqnarray}
  \frac{\Omega}{2\pi ce} j_{\rm gap} 
  &\equiv& \frac{N_+(s_2)}{B(s_2)}-\frac{N_+(s_1)}{B(s_1)}
  \nonumber\\
  &=&      \frac{N_-(s_1)}{B(s_1)}-\frac{N_-(s_2)}{B(s_2)},
  \label{eq:def_jgap}
\end{eqnarray}
where $s_1$ and $s_2$ designate the position of the
inner and the outer boundaries, respectively.
That is, $W= s_2 -s_1$.
Equation~(\ref{eq:closure}) corresponds to a generalized verion of the 
gap closure condition considered in
Papers~IV and V (e.g. eq.~[30] in Paper~V), 
in which $j_1=j_2=0$ and hence $j_{\rm gap}= j_{\rm tot}$ was assumed.
When there is a current injection 
(i.e., when $j_1$ or $j_2$ is non-vanishing),
not only the produced particles in the gap
but also the injected particles contribute for the $\gamma$-ray emission.
Therefore, the gap width is adjusted smaller compared with 
$j_1=j_2=0$ case by the factor $j_{\rm gap}/j_{\rm tot}$.
Utilizing condition~(\ref{eq:closure}),
we can rewrite equation~(\ref{eq:master_eq_1}) into the form
\begin{equation} 
  \frac{d^2}{ds^2} \left(\frac{N_+ -N_-}{B}\right)
  = 4 \frac{j_{\rm tot}}{j_{\rm gap}}
      \frac{1}{W} \frac{N_+ -N_-}{B}.
  \label{eq:master_eq}
\end{equation}

To solve the differential equation~(\ref{eq:master_eq}),
we impose the following two bounday conditions:
\begin{equation}
  ce \frac{N_+(s_1)}{B(s_1)}= \frac{\Omega}{2\pi} j_1,
  \label{eq:BD_extra_1}
\end{equation}
\begin{equation}
  ce \frac{N_-(s_2)}{B(s_2)}= \frac{\Omega}{2\pi} j_2.
  \label{eq:BD_extra_2}
\end{equation}
With the aid of equation~(\ref{eq:def_jgap}),
these two bounday conditions give
\begin{equation}
  \frac{N_+ -N_-}{B}= -\frac{\Omega}{2\pi}(j_{\rm gap}-j_1+j_2)
  \label{BD_01}
\end{equation}
at $s= s_1$, and 
\begin{equation}
  \frac{N_+ -N_-}{B}= \frac{\Omega}{2\pi}(j_{\rm gap}+j_1-j_2)
  \label{BD_02}
\end{equation}
at $s= s_2$.
It follows from equation~(\ref{eq:master_eq}) 
\begin{eqnarray}
  \frac{N_+ -N_-}{B}
  &=& \frac{\Omega}{2\pi ce}
      \left[ j_{\rm gap} 
             \frac{\sinh\left(\sqrt{\displaystyle\frac{j_{\rm tot}}
                                                      {j_{\rm gap}}
                                   }
                               \displaystyle\frac{s-s_{\rm cnt}}
                                                {W/2}
                        \right)
                  }
                  {\sinh\left(\sqrt{\displaystyle\frac{j_{\rm tot}}
                                                      {j_{\rm gap}}
                                   }
                        \right)
                  }
      \right.
  \nonumber\\
  & &
  \hspace*{-2.0 truecm}
  + \left. (j_1-j_2)
             \frac{\cosh\left(\sqrt{\displaystyle\frac{j_{\rm tot}}
                                                      {j_{\rm gap}}
                                   }
                               \displaystyle\frac{s-s_{\rm cnt}}
                                                {W/2}
                        \right)
                  }
                  {\cosh\left(\sqrt{\displaystyle\frac{j_{\rm tot}}
                                                      {j_{\rm gap}}
                                   }
                        \right)
                  }
      \right],
  \label{eq:charge_density}
\end{eqnarray}
where the gap center position is defined by
\begin{equation}
  s_{\rm cnt} \equiv \frac{s_2-s_1}{2}.
  \label{eq:def_scnt}
\end{equation}

\subsection{Poisson Equation}
\label{sec:Poisson}

The real charge density $e(N_+ -N_-)$, which is given by 
equation~(\ref{eq:charge_density}) appears in the Poisson equation
for the non-corotational potential $\Psi$. 
Neglecting relativistic effects,
and assuming that typical transfield thickness of the gap, $D_\perp$, 
is greater than or comparable with $W$,
we can reduce the Poisson equation into the one-dimensional form 
(Paper~VI; see also \S~2 in Michel 1974)
\begin{equation}
 -\nabla^2 \Psi 
    = 4\pi \left[ e(N_+ -N_-) +\frac{\Omega B_z}{2\pi c} \right],
  \label{eq:Poisson_0}
\end{equation}
where $e$ designates the magnitude of the charge on an electron.
Substituting equation~(\ref{eq:charge_density}) into 
(\ref{eq:Poisson_0}), we obtain
\begin{eqnarray}
  -\nabla^2 \Psi 
    &=& \frac{2B\Omega}{c}
        \left[ j_{\rm gap} 
                  f_{\rm odd} \left( \frac{s-s_{\rm cnt}}{W/2} \right)
        \right.
    \nonumber\\
    & & \hspace*{-1.0 truecm}
        \left.
              +(j_1-j_2)
                  f_{\rm even}\left( \frac{s-s_{\rm cnt}}{W/2} \right)
              +\frac{B_z}{B}
        \right],
  \label{eq:Poisson_1}
\end{eqnarray}
where 
\begin{equation}
  f_{\rm odd}(x) 
  \equiv \frac{\sinh\left(\sqrt{\displaystyle\frac{j_{\rm tot}}
                                                  {j_{\rm gap}}
                               }
                          \, x     
                    \right)
              }
              {\sinh\left(\sqrt{\displaystyle\frac{j_{\rm tot}}
                                                  {j_{\rm gap}}
                               }
                    \right)
              }
  \label{eq:Poisson_1a}
\end{equation}
and
\begin{equation}
  f_{\rm even}(x) 
  \equiv \frac{\cosh\left(\sqrt{\displaystyle\frac{j_{\rm tot}}
                                                  {j_{\rm gap}}
                               }
                          \, x     
                    \right)
              }
              {\cosh\left(\sqrt{\displaystyle\frac{j_{\rm tot}}
                                                  {j_{\rm gap}}
                               }
                    \right)
              }.
  \label{eq:Poisson_1b}
\end{equation}
There are essentially three assumptions that are used to derive
equations~(\ref{eq:Poisson_1}), (\ref{eq:Poisson_1a}), 
and (\ref{eq:Poisson_1b}):
the radiation-reaction forces exactly cancel with the 
electrostatic force in the particles' Boltzmann equations;
$\eta_{{\rm p}+}(\Eg)=\eta_{{\rm p}-}(\Eg)$,
which may be justified for a power-law, magnetospheric X-ray component;
and the Poisson equation is analyzed one-dimensionally along the
magnetic field line.

\subsection{Generalization of the Null Surface}
\label{sec:null}

To examine the Poisson equation~(\ref{eq:Poisson_1}) analytically,
we assume that the transfield thickness of the gap is 
greater than $W$ and replace $\nabla^2 \Psi$ with $d^2 \Psi/ds^2$.
Furthermore, we neglect the current created in the gap
and simply set $j_{\rm gap}=0$.

First, consider the case when a current injects from neither of the
boundaries, that is, $j_1=j_2=0$.
It follows that the
derivative of the acceleration field (i.e., $-d^2\Psi/ds^2$) vanishes
at the conventional null surface where $B_z$ vanishes.
We may notice that
$-d^2\Psi/ds^2$ is positive at the inner part of the gap
and changes its sign near the gap center ($s= s_{\rm cnt}$) 
to become negative at the outer part of the gap. 
Therefore, we can conclude that the gap is located 
(or centers) around the conventional null surface,
if there is no current injection from outside.

Secondly, consider the case when a current is injected at the
inner boundary (at $s=s_1$) and $j_1-j_2>0$ holds.
Since the function $f_{\rm even}$ is positive at arbitrary $s$,
the gap center is located at a place where $B_z$ is negative,
that is, outside of the conventional null surface. 
In particular, when $j_1-j_2 \sim 1$ holds, 
$-d^2\Psi/ds^2$ vanishes at the place where $B_z \sim -B$.
In a vacuum, static dipole field,
$B_z \sim -B$ is realized along the last-open field line
at the light cylinder.
Therefore, the gap is expected to shift towards the light cylinder,
if the injected current density at the inner boundary
approaches the Goldreich-Julian value. 
We may notice here that $f_{\rm even}$ is less than unity,
because $\vert s-s_{\rm cnt} \vert$ does not exceed $W/2$.

Thirdly and finally, consider the case when $j_1-j_2 \sim -1$ holds.
In this case, $-d^2\Psi/ds^2$ vanishes at the place where $B_z \sim B$.
Therefore, gap is expected to be located close to the star surface,
if a Goldreich-Julian current density is injected at the outer boundary. 
In what follows, we will examine more accurately these predictions 
on the gap position vs. current injection,
by solving the Vlasov equations~(\ref{eq:cont-eq}), 
(\ref{eq:Boltz_gam}), and (\ref{eq:Poisson_1}) numerically.

\section{Basic Equations and Boundary Conditions}
\label{sec:reduction}

In the present paper, we assume that the transfield thickness, $D_\perp$,
of the gap is much greater than $W$,
and neglect the transfield derivatives in the 
Poisson equation~(\ref{eq:Poisson_0}).
We consider that this one-dimensional analysis could be justified
because $D_\perp \sim 6 W$ is required so that 
the predicted GeV flux may be consistent with the EGRET observations
(\S~\ref{sec:res_Ell}).
We rewrite the Vlasov equations
into the suitable forms for numerical analysis
in \S~\ref{sec:basic}, 
and impose boundary conditions 
in \S~\ref{sec:BD}.

\subsection{One-dimensional Vlasov Equations}
\label{sec:basic}

As will be shown at the end of this section,
it is convenient to introduce the 
typical Debey scale length $c/\omega_{\rm p}$, 
\begin{equation}
  \omega_{\rm p} = \sqrt{ \frac{4\pi e^2}{m_{\rm e}}
	                  \frac{\Omega \Bc}{2\pi ce} },
  \label{eq:def-omegap}
\end{equation}
where $\Bc$ represents the magnetic field strength at the gap center.
The dimensionless coordinate variable then becomes
\begin{equation}
  \xi \equiv (\omega_{\rm p}/c) s.
  \label{eq:def-xi}
\end{equation}
By using such dimensionless quantities, we can rewrite
the Poisson equation into
\begin{equation}
  E_\parallel = -\frac{d\psi}{d\xi},
  \label{eq:basic-1}
\end{equation}
\begin{equation}
  \frac{dE_\parallel}{d\xi}
  = \frac{B(\xi)}{\Bc} \left[ n_+(\xi) - n_-(\xi) \right]
    + \frac{B_z(\xi)}{\Bc}
  \label{eq:basic-2}
\end{equation}
where $ \psi(\xi) \equiv e\Psi(s)/(m_{\rm e}c^2)$;
the particle densities per unit flux tube are defined by
\begin{equation}
  n_\pm(\xi) \equiv 
    \frac{2\pi ce}{\Omega} \frac{N_\pm}{B}.
  \label{eq:def-n}
\end{equation}
We evaluate $B_z/B$ at each point along the last-open field line,
by using the Newtonian dipole field.

Let us introduce the following dimensionless $\gamma$-ray 
densities in the dimensionless energy interval
between $\beta_{i-1}$ and $\beta_i$:
\begin{equation}
  g_\pm^i(\xi) \equiv 
    \frac{2\pi ce}{\Omega \Bc}
    \int_{\beta_{i-1}}^{\beta_i} d\epsilon_\gamma G_\pm(s,\epsilon_\gamma).
  \label{eq:def-g}
\end{equation}
In this paper, we set $\beta_0=10^2$,
which corresponds to the lowest $\gamma$-ray energy, $51.1$ MeV.
We divide the $\gamma$-ray spectra into $9$ energy bins 
and put
$\beta_1= 10^{2.5}$, 
$\beta_2= 10^3$, 
$\beta_3= 10^{3.5}$, 
$\beta_4= 10^4$, 
$\beta_5= 10^{4.5}$, 
$\beta_6= 10^{4.75}$, 
$\beta_7= 10^5$. 
$\beta_8= 10^{5.25}$, and
$\beta_9= 10^{5.5}$.

We can now rewrite the continuity quation~(\ref{eq:cont-eq}) 
of particles into 
\begin{equation}
  \frac{dn_\pm}{d\xi} = 
    \pm \frac{\Bc}{B\cos\Phi}
        \sum_{i=1}^{9} [ \eta_{\rm p+}{}^i g_+^i(\xi)
                        +\eta_{\rm p-}{}^i g_-^i(\xi)],
  \label{eq:basic-3}
\end{equation}
where the magnetic field strength, $B$, is evaluated at each $\xi$.
The dimensionless redistribution functions
$\eta_{{\rm p}\pm}^i$ 
are evaluated at the central energy in each bin as
\begin{equation}
  \eta_{{\rm p}\pm}^i \equiv
  \frac{1}{\omega_{\rm p}}
  \eta_{{\rm p}\pm}\left(\frac{\beta_{\rm i-1}+\beta_{\rm i}}{2}\right).
  \label{eq:def_etap_1}
\end{equation} 

A combination of equations (\ref{eq:basic-3}) 
gives the current conservation law,
\begin{equation}
  j_{\rm tot} \equiv n_+(\xi) + n_-(\xi) = {\rm constant \ for \ } \xi,
  \label{eq:consv}
\end{equation}
which is equivalent with equation~(\ref{eq:consv_0}).

The Boltzmann equations~(\ref{eq:Boltz_gam}) for the $\gamma$-rays
are integrated over $\Eg$ between
dimensionless energies $\beta_{\rm i-1}$ and $\beta_{\rm i}$ 
to become
\begin{eqnarray}   
  \frac{d}{d\xi} g_\pm^i
     = \frac{d}{d\xi}\left( \ln B \right)
       \mp \frac{\eta_{{\rm p}\pm}{}^i}{\cos\Phi} g_\pm^i
       \pm \frac{\eta_{\rm c}^i B(\xi)}{\Bc\cos\Phi} n_\pm,
  \label{eq:basic-5}
\end{eqnarray}   
where $i=1,2,\cdot\cdot\cdot,m$ ($m=9$) and 
\begin{eqnarray}
  \eta_{\rm c}^i 
  &\equiv& \frac{\sqrt{3}e^2\Gamma}{\omega_{\rm p}hR_{\rm c}}
           \int_{\beta_{i-1} / \epsilon_{\rm c}}
               ^{\beta_i     / \epsilon_{\rm c}}
            ds \int_s^\infty K_{\frac53}(t)dt
  \label{eq:etaCi}
\end{eqnarray}
is dimensionless.

Equating the electric force $e \vert d\Psi / dx \vert$ and the
radiation reaction force,
we obtain the saturated Lorentz factor at each point as 
\begin{equation}
  \Gamma_{\rm sat} 
   = \left( \frac{3 R_{\rm c}{}^2}{2e} 
		    \left| \frac{d\Psi}{ds} \right|
                  + 1 
     \right)^{1/4};
  \label{eq:saturated}
\end{equation}
we compute the curvature radius $R_{\rm c}$ 
at each point for a Newtonian dipole magnetic field.
Since the maximum of $\vert d\Psi/dx \vert$ and the potential drop
are roughly proportional to $W^2$ and $W^3$, respectively (Paper~V),
the particles become unsaturated for very small $W$.
To avoid an overestimation of the Lorentz factor in such cases, 
we compute $\Gamma$ by 
\begin{equation}
  \frac{1}{\Gamma}
  = \sqrt{ \frac{1}{\Gamma_{\rm sat}{}^2}
          +\frac{1}{\psi^2(\xi_2)}
         },
  \label{eq:terminal}
\end{equation}
where $\psi(\xi_2)$ represents the maximum attainable Lorentz factor.

\subsection{Boundary Conditions}
\label{sec:BD}

We now consider the boundary conditions 
to solve the Vlasov equations
(\ref{eq:basic-1}), (\ref{eq:basic-2}), (\ref{eq:basic-3}), 
and (\ref{eq:basic-5}).
At the {\it inner} (starward) boundary
($\xi= \xi_1$), we impose (Paper~VI)
\begin{equation}
  E_\parallel(\xi_1)=0,
  \label{eq:BD-1}
\end{equation}
\begin{equation}
  \psi(\xi_1) = 0,
  \label{eq:BD-2}
\end{equation}
\begin{equation}
  g_+^i(\xi_1)=0  \quad (i=1,2,\cdot\cdot\cdot,9).
  \label{eq:BD-3}
\end{equation}
It is noteworthy that condition (\ref{eq:BD-1}) is consistent with
the stability condition at the plasma-vacuum interface 
if the electrically supported magnetospheric plasma
is completely-charge-separated, 
i.e., if the plasma cloud at $\xi < \xi_1$ is composed of 
electrons alone (Krause-Polstorff \& Michel 1985a,b; Michel 1991).
We assume that the Goldreich-Julian plasma gap boundary 
is stable with $\Ell=0$ on the boundary, $\xi=\xi_1$.

Since positrons may flow into the gap at $\xi=\xi_1$
as a part of the global current pattern in the magnetosphere,
we denote the positronic current per unit flux tube at $\xi=\xi_1$ as
\begin{equation}
  n_+(\xi_1)= j_1,
  \label{eq:BD-4}
\end{equation}
which yields (eq.~[\ref{eq:consv}])
\begin{equation}
  n_-(\xi_1)= j_{\rm tot}-j_1.
  \label{eq:BD-5}
\end{equation}

At the {\it outer} boundary ($\xi=\xi_2$), we impose
\begin{equation}
  E_\parallel(\xi_2)=0,
  \label{eq:BD-6}
\end{equation}
\begin{equation}
  g_-^i(\xi_2)=0 \quad (i=1,2,\cdot\cdot\cdot,9),
  \label{eq:BD-7}
\end{equation}
\begin{equation}
  n_-(\xi_2)= j_2.
  \label{eq:BD-8}
\end{equation}
Conditions~(\ref{eq:BD-4}) and (\ref{eq:BD-8}) are equivalent with
(\ref{eq:BD_extra_1}) and (\ref{eq:BD_extra_2}). 

The current density created in the gap per unit flux tube
can be expressed as
\begin{equation}
  j_{\rm gap}= j_{\rm tot} -j_1 -j_2.
  \label{eq:Jgap}
\end{equation}
This equation is, of course, consistent with equation~(\ref{eq:def_jgap}).
We adopt $j_{\rm gap}$, $j_1$, and $j_2$
as the free parameters.

We have totally $24$ boundary conditions 
(\ref{eq:BD-1})--(\ref{eq:BD-8})
for $22$ unknown functions
$\Psi$, $E_\parallel$,
$n_\pm$, 
$g_\pm^i$ ($i=1, 2, \cdot\cdot\cdot, 9$).
Thus two extra boundary conditions must be compensated 
by making the positions of the boundaries $\xi_1$ and $\xi_2$ be free.
The two free boundaries appear because $E_\parallel=0$ is imposed at 
{\it both} the boundaries and because $j_{\rm gap}$ is externally imposed.
In other words, the gap boundaries ($\xi_1$ and $\xi_2$) shift,
if $j_1$ and/or $j_2$ varies.

Let us briefly comment the convenience of the introduction of the
dimensionless coordinates and variables.
It follows from the Vlasov equations~(\ref{eq:basic-1}),
(\ref{eq:basic-2}), (\ref{eq:basic-3}), and (\ref{eq:basic-5})
that the solutions $\xi$, $\psi$, $\Ell$, $n_\pm$, and $g_\pm^i$
are unchanged if $B$, $\eta_{{\rm p}\pm}^i$, and $\eta_{\rm c}^i$
are invariant.
Consider the case when the normalization of $dN_{\rm x}/d\Ex$ 
is doubled.
In this case, equations~(\ref{eq:def_etap_0}) and (\ref{eq:def_etap_1})
show that $\eta_{\rm p}^i$ is invariant
if we also double $\omega_{\rm p}$. 
Note that $\epsilon_{\rm c}$ in equation~(\ref{eq:etaCi}) is 
proportional to $\omega_{\rm p}^{3/4} R_{\rm c}^{1/2}$
(eq.~[\ref{eq:def_Ec}]),
where 
$\Gamma \sim \Gamma_{\rm sat} 
        \propto R_{\rm c}^{1/2} \vert d\Psi/ds \vert^{1/4}$
is used.
It follows that $\epsilon_{\rm c}$ is also invariant 
if we increase $R_{\rm c}$ by $2^{-3/2}$ times.
It should be noted that $\eta_{\rm c}^i$ (eq.~[\ref{eq:etaCi}]) is 
invariant by this change of parameters.
On these grounds, we can reduce one degree of freedom in the
free parameters.

\section{Predicted Gamma-ray Flux}
\label{sec:Gamma-ray_spc}

In this section, we detail the method how to compute 
$\nu F_\nu$ spectrum in GeV energies in \S~\ref{sec:GeV_spc}
and in TeV energies in \S~\ref{sec:TeV_spc}.

\subsection{GeV Spectra}
\label{sec:GeV_spc}

The GeV spectra of outwardly and inwardly propagating $\gamma$-rays
are obtained from $g_+^i(\xi_2)$ and $g_-^i(\xi_1)$. 
At position $\xi$, the $\gamma$-ray emission rate becomes
\begin{equation}
  \mbox{$\gamma$-ray flux}=
  A_{\rm cr}(\xi) \, c \, \frac{\Omega\Bc}{2\pi ce} g_\pm^i(\xi)
  \, \mbox{s}^{-1} ,
  \label{eq:gamma_flux}
\end{equation}
where $A_{\rm cr}$ refers to the cross section of the gap at $\xi$.
Multiplying the mean $\gamma$-ray energy 
$\sqrt{\beta_{\rm i}\beta_{\rm i-1}} m_{\rm e}c^2$,
on equation~(\ref{eq:gamma_flux}), 
dividing it by $\Delta \Omega_{\rm GeV} d^2$,
and further dividing by the frequency interval
$m_{\rm e}c^2(\beta_{\rm i}-\beta_{\rm i-1})/h$,
we obtain the flux density, $F_\nu$;
here, $\Delta \Omega_{\rm GeV}$ is the emission solid angle,
and $h$ the Planck constant.
We thus obtain the GeV flux
\begin{equation}
  \nu F_\nu = \frac{\beta_{\rm i} \beta_{\rm i-1}}
                   {\beta_{\rm i}-\beta_{\rm i-1}}
              m_{\rm e}c^2
              \frac{\Omega\Bc}{2 \pi e}
              \frac{A_{\rm cr} g_\pm^{\rm i}}
                   {\Delta \Omega_{\rm GeV} \, d^2}.
  \label{eq:nuFnu_GeV}
\end{equation}
To compute the $\gamma$-ray flux emitted outwardly (or inwardly) 
from the gap,
we adopt the plus (or the minus) sign in $g_\pm$
and evaluate $A_{\rm cr}g_+$ at $\xi= \xi_2$ (or $\xi_1$).
As will be shown in \S~\ref{sec:results}, 
$W \ll \rlc$ holds for the Crab pulsar.
We thus simply apply the same cross section for both the 
outwardly and inwardly emitted $\gamma$-rays and put
$A_{\rm cr}= D_\perp^2$,
where $D_\perp$ should be greater than or at least comparable with $W$
for the one-dimensional approximation of the Poisson equation 
(\S~\ref{eq:Poisson_0}) to be justified.

It is noteworthy that the particles lose most of their energy
in the gap 
if $l_{\rm acc} \ll W$ holds,
where $l_{\rm acc}$ refers to the length scale for particles
to be accelerated to the saturated Lorentz factor 
(eq.~[\ref{eq:saturated}]).
That is, we can neglect the primary luminosity emitted by 
the particles running {\it outside} of the gap,
compared with that emitted by the particles running
{\it inside} of the gap, if $l_{\rm acc} \ll W$.
Since the mono-energetic approximation of the particle motion
(\S~\ref{sec:cont}) is justified when $l_{\rm acc} \ll W$,
the neglect of GeV emission by the particles running outside of the gap
is consistent with the mono-enegetic approximation.
We thus compute the GeV luminosity 
from the solved $\gamma$-ray distribution functions
$g_+^i(\xi_2)$ and $g_-^i(\xi_1)$.
 
\subsection{TeV Spectra}
\label{sec:TeV_spc}

Once the electrodynamic structure of the gap is solved,
we can further compute the upscattered $\gamma$-ray flux
emitted from the whole accelerator,
if additionally give the infrared photon field.
This treatment is justified unless 
the upscattered, TeV luminosity exceeds 
the curvature-radiated, GeV one.

If an electron or a positron is migrating 
with Lorentz factor $\Gamma \gg 1$ in an isotropic photon field,
it upscatters the soft photons to produce
the following number spectrum of $\gamma$-rays
(Blumenthal \& Gould 1970):
\begin{eqnarray}
  \frac{d N}{dtd\Eg}
  &=& \frac34 \sgT \frac{c}{\Gamma^2}
      \frac{dN_{\rm IR}}{d\epsilon_{\rm IR}}
      \frac{d\epsilon_{\rm IR}}{\epsilon_{\rm IR}}
  \nonumber \\
  & & \hspace{-2.0 truecm}
      \times
      \left[ 2q \ln q +(1+2q)(1-q)
            +\frac{(Qq)^2(1-q)}{2(1+Qq)}
      \right],
  \label{eq:spc_tev}
\end{eqnarray}
where $Q \equiv 4 \epsilon_{\rm IR} \Gamma$,
$q \equiv \Eg / Q(\Gamma-\Eg)$,
$dN_{\rm IR} / d\epsilon_{\rm IR}$ refers to the IR photon density
per unit dimensionless enegy interval between 
$\epsilon_{\rm IR}$ and $\epsilon_{\rm IR}+d\epsilon_{\rm IR}$,
$\sgT$ is the Thomson cross section;
$\epsilon_{\rm IR}$ and $\Eg$ are the energies of the IR and the 
upscattered photons in units of $m_{\rm e}c^2$.
Equation~(\ref{eq:spc_tev}) is valid if the resonance effects are
negligible, that is, $B \ll B_{\rm crit} = 4.4 \times 10^{13}$~G.
This inequality is satisfied except for the polar cap.
The flux density of the upscattered photons becomes
\begin{equation}
  F_\nu= \frac{N_{\rm e}}{\Delta \Omega_{\rm TeV} \, d^2} \cdot
         h\Eg \int_{\epsilon_{\rm IR,min}}^{\epsilon_{\rm IR,max}}
                 \frac{d N}{dt d\Eg d\epsilon_{\rm IR}} 
                 d\epsilon_{\rm IR},
  \label{eq:Fnu_TeV}
\end{equation}
where $\Delta \Omega_{\rm TeV}$ refers to the emission solid angle
of the upscattered photons.
In this paper, we estimate $N_{\rm e}$ with
\begin{equation}
  N_{\rm e}= (j_{\rm gap}+j_1)
             \frac{\Omega\Bc}{2\pi ce} W D_\perp{}^2
  \label{eq:Ne_1}
\end{equation}
to compute the outwardly propagating TeV flux, 
which are emitted by outwardly propagating particles (i.e., positrons) 
and with
\begin{equation}
  N_{\rm e}= (j_{\rm gap}+j_2)
             \frac{\Omega\Bc}{2\pi ce} W D_\perp{}^2
  \label{eq:Ne_2}
\end{equation}
to compute the inwardly propagating TeV flux,
which are emitted by inwardly propagating particles (i.e., electrons). 

Multiplying the $\gamma$-ray frequency $\Eg m_{\rm e}c^2/h$
on the $F_\nu$ flux density (eq.~[\ref{eq:Fnu_TeV}]), 
we obtain the upscattered flux 
\begin{eqnarray}
   \nu F_\nu 
   &=& (j_{\rm gap}+j_{\rm a})
               m_{\rm e}c^2
               \frac{\Omega\Bc}{2 \pi e}
               \frac{A_{\rm cr}}{\Delta \Omega_{\rm TeV} d^2}
               \Eg{}^2 
   \nonumber\\
   & & \times W
               \int_{\epsilon_{\rm IR,min}}^{\epsilon_{\rm IR,max}}
                  \frac{d N}{cdt d\Eg d\epsilon_{\rm IR}} 
                  d\epsilon_{\rm IR},
  \label{eq:nuFnu_TeV}
\end{eqnarray}
where $j_{\rm a}= j_1$ (or $j_2$) for outwardly (or inwardly)
emitted $\gamma$-rays.
As the emission solid angles,
we assume 
\begin{equation}
  \Delta\Omega_{\rm GeV}
  = \Delta\Omega_{\rm TeV}
  = \frac{2 \pi W}{\rlc}
  \label{eq:solid_ang}
\end{equation}
in this paper.

We also consider the extrinsic absorption of the TeV photons
outside of the gap.
For a homogeneous and isotropic IR field,
the optical depth becomes
\begin{equation}
  \tau(\Eg) 
  = \frac{\rlc}{2}
    \int_{\epsilon_{\rm IR,min}}^{\epsilon_{\rm IR,max}}
          \frac{dN_{\rm IR}}{d\epsilon_{\rm IR}}
          \sgP(\epsilon_{\rm IR},\Eg,\mu_{\rm c}) d\epsilon_{\rm IR},
  \label{eq:tauTeV}
\end{equation}
where the path length is assumed to be $\rlc/2$.
We apply the same path length for all the cases considered
so that the extrinsic absorption may work equally.

\section{Application to the Crab Pulsar}
\label{sec:app_Crab}

\subsection{Input Infrared Field}
\label{sec:IR}

Consider the case when the IR spectrum is homogeneous and
expressed by a single power-law,
\begin{equation}
  \frac{dN_{\rm IR}}{d\epsilon_{\rm IR}}
    = N_0 \epsilon_{\rm IR}^\alpha,
  \label{eq:IR_spc}
\end{equation}
where $N_0$ and $\alpha$ are spatially constant.
For an isotropic field, the specific intensity becomes
\begin{equation}
  I_\nu = \frac{c}{4\pi} h N_0 \epsilon_{\rm IR}^{\alpha+1}.
  \label{eq:IR_intensity}
\end{equation}
Assuming that this uniform sphere has radius $\rlc$,
we obtain the following flux density at distance $d$:
\begin{eqnarray}
  F_\nu 
  &=& \frac{c}{4} \left(\frac{\rlc}{d}\right)^2 h N_0
      \epsilon_{\rm IR}{}^{\alpha+1}
  \nonumber\\
  & & \hspace*{-0.5 truecm}
  = 4.5 \times 10^{-20} \Omega_2{}^{-2}
      \left(\frac{d}{\rm kpc}\right)^{-2}
      N_0 \epsilon_{\rm IR}{}^{\alpha+1}
      \mbox{Jy}.
  \label{eq:flux_density}
\end{eqnarray}
As the lower and upper cutoff IR photon energies, 
we adopt 
$\epsilon_{\rm IR,min}= 10^{-8}$ and
$\epsilon_{\rm IR,max}= 10^{-6}$,
where $\epsilon_{\rm IR,min} < \epsilon_{\rm IR} < \epsilon_{\rm IR,max}$.

\begin{figure} 
\centerline{ \epsfxsize=8.5cm \epsfbox[200 2 700 350]{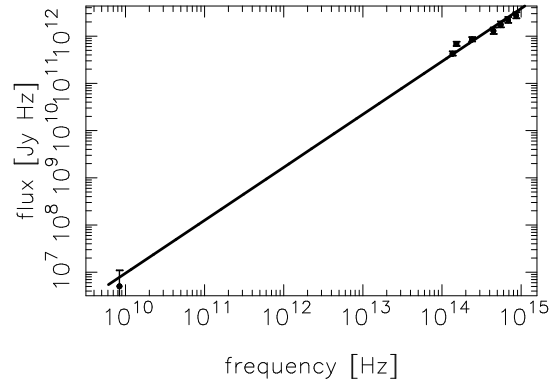} } 
\caption{\label{fig:spc_Crab} 
A single power-law fit of phase-averaged color spectrum
of the Crab pulsar (case~B, see \S~5.1).
The abscissa is the photon frequency in Hz,
while the ordinate is the photon flux in Jy~Hz.
        }
\end{figure}

Because the pulsed flux around eV energies are difficult to 
be observed,
we consider the following two cases for the set of $N_0$ and $\alpha$:\\
{\bf case A} \quad
We assume that the IR spectrum below 
$\epsilon_{\rm IR} < 10^{-6}$ 
(or equivalently, below $1.23 \times 10^{14}$~Hz)
are optically thick for synchrotron self-absorption
and adopt $\alpha= 1.5$.
Setting $F_\nu=3$~mJy at $\epsilon_{\rm IR} = 10^{-6}$, 
which is consistent with near-IR and optical observations 
(Eikenberry et al. 1997),
we obtain $N_0= 2.3 \times 10^{32} \mbox{cm}^{-3}$.\\
{\bf case B} \quad
Interpolating the phase-averaged color spectrum in 
UV, U, B, V, R (Percival et al. 1993),
J, H, K (Eikenberry et al. 1997) bands,
and the radio observation at 8.4 GHz (Moffett and Hankins 1996),
we obtain $N_0= 1.5 \times 10^{17} \mbox{cm}^{-3}$ and $\alpha=-0.88$.
In figure~\ref{fig:spc_Crab}, we present the fitted spectrum with the
solid line;
the ordinate is $\nu F_\nu$ in Jy~Hz.

\subsection{Input X-ray Field}
\label{sec:X}

HEAO 1 observations revealed that the
X-ray spectrum in the primary pulse phase 
is expressed by 
\begin{equation}
  \frac{dN_{\rm pl}}{d\epsilon_{\rm x}}
    = N_{\rm pl} \epsilon_{\rm x}{}^\alpha
  \quad (\epsilon_{\rm min} < \epsilon_{\rm x} < \epsilon_{\rm max}),
  \label{eq:dNdE_0}
\end{equation}
with $\alpha=-1.81$ 
and $N_{\rm pl}=5.3 \times 10^{15} (d/{\rm kpc})^2 (r_{\rm cnt}/\rlc)^{-2}$
(Knight 1982).
We adopt
$\epsilon_{\rm min}=0.1 \mbox{keV} / 511 \mbox{keV}$ and
$\epsilon_{\rm max}= 50 \mbox{keV} / 511 \mbox{keV}$.
Unlike the IR field, which is assumed to be homogeneous within radius 
$\rlc$,
we suppose that the X-rays are emitted near to the gap.
In this case, the X-ray density computed from the observed flux will 
increase as the gap is located close to the star.
To consider such effects,
we simply assume that the X-ray density is proportional to the
inverse square of $r_{\rm cnt}$.

The angle dependence of the specific intentity of the
X-ray field is considered in the collision angle, $\mu_{\rm c}$ 
(eq.~[\ref{eq:def_etap_0}], or [\ref{eq:def_etap_1}]).
In the case of the Crab pulsar, 
the X-ray field is dominated by a power-law component,
which is probably emitted near the outer-gap accelerator
rather than from the neutron star surface.
We thus simply evaluate the cosine of the collision angles as
\begin{equation} 
  \mu_{\rm c}= \cos(W/\rlc),
  \label{eq:def_coll_power}
\end{equation}
for both inwardly and outwardly propagating $\gamma$-rays.
Aberration of light is not important for this component,
because both the X-rays and the $\gamma$-rays are emitted
nearly at the same place.
We may notice here that this is a rough estimate of $\mu_{\rm c}$
and that $\epsilon_{\rm th}= 2/[(1-\mu_{\rm c})\epsilon_\gamma]$
strongly depends on $\mu_{\rm c}$ if $W \ll \rlc$
(i.e., if $1-\mu_{\rm c} \ll 1$).
In the case of the Crab pulsar, $W/\rlc \sim 0.05$ holds
(see fig.~2);
therefore, the true results will depend on the detailed beaming geometry
of the secondary X-rays,
which are emitted outside of the gap along local magnetic field lines
via synchrotron process.
However, to inquire into this matter would lead us to into that 
specialized area of the magnetic field configuration
close to the light cylinder.
Such a digression would undoubtedly obscure the outline of our 
argument.

\subsection{Results}
\label{sec:results}

Let us now substitute the X-ray field into
equation~(\ref{eq:def_etap_0}) and 
solve the Vlasov equations by the method described in \S~2.
It should be noted that 
we do not use the IR spectrum considered in
\S~\ref{sec:IR} to solve $\Ell(\xi)$, $n_\pm(\xi)$, and $g_\pm{}^i(\xi)$;
they are necessary when we consider the TeV emission due to IC scatterings.

In Paper~I, it is argued that the IC scatterings dominates
the curvature radiation
in the outer gap in the Crab pulsar magnetosphere if we
adopt $\mu_{\rm c}=0$.
However, in the present paper, we consider that
the collision angles are much smaller than $90^\circ$
and adopt $\mu_{\rm c}= \cos(W/\rlc)$.
In this case, unless the gap is located well inside of the
conventional null surface,
curvature radiation becomes the primary process in $\gamma$-ray
emission, and hence in the radiation-reaction forces.
The rotational frequency and the magnetic moment are
$\Omega= 188.1 \mbox{rad s}^{-1}$ and 
$\mu_{\rm m}= 3.38 \times 10^{30} \mbox{G cm}^3$.

\subsubsection{Electric Field Structure}
\label{sec:res_Ell}

To reveal the spatial distribution of the acceleration field, 
we consider four representative boundary conditions:\\
{\bf case~1} \quad  $(j_1,j_2)=(0,0)$   $\rightarrow$ solid curves, \\
{\bf case~2} \quad  $(j_1,j_2)=(0.3,0)$ $\rightarrow$ dashed curves, \\ 
{\bf case~3} \quad  $(j_1,j_2)=(0.6,0)$ $\rightarrow$ dash-dotted curves, \\
{\bf case~4} \quad  $(j_1,j_2)=(0,0.3)$ $\rightarrow$ dotted curves. \\
That is, for case~2 (or case~4), 
the positronic (or electronic) current density flowing into the gap
per unit flux tube at the inner (or outer) boundary is $30\%$ of 
the typical Goldreich-Julian value, $\Omega /2\pi$.
We fix $j_{\rm gap}= 0.01$ for all the four cases,
because the solution forms a \lq brim' 
to disappear (fig.~2 in Hirotani \& Okamoto 1998)
if $j_{\rm gap}$ exceeds a few percent.
In what follows, we adopt $45^\circ$ as the magnetic inclination,
which is necessary to compute $B$ at each point 
for the Newtonian dipole field.
Since the X-ray field is dominated by a power-law component,
the inclination does not affect $\mu_{\rm c}$.

\begin{figure} 
\centerline{ \epsfxsize=8.5cm \epsfbox[200 20 500 250]{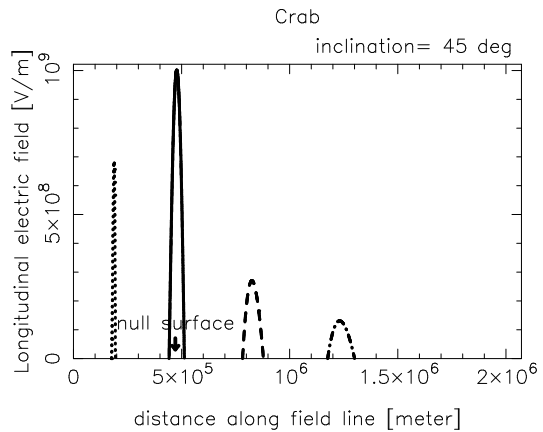} } 
\caption{\label{fig:Ell_Crab_45} 
Distribution of $\Ell(s)$ for the Crab pulsar with $\inc=45^\circ$; 
the abscissa is in meters.
The solid, dashed, dash-dotted, and dotted curves correspond to the
cases 1, 2, 3, and 4, respectively (see text).
        }
\end{figure}

The results of $\Ell(\xi)$ for the four cases are presented in
figure~\ref{fig:Ell_Crab_45}.
The abscissa designates the distance along the last-open field line
and covers the range from the neutron star surface ($s=0$)
to the position where the disance equals 
$s= 1.2 \rlc= 1.91 \times 10^6$~m.

The solid curve (case~1) shows that the gap is located around the 
conventional null surface. 
However, the gap shifts outwards as $j_1$ increases,
as the dashed (case~2) and dash-dotted (case~3) curves indicate.
This result is consistent with what was predicted in 
Shibata and Hirotani (2000) analytically.

On the other hand, when $j_2$ increases,
the gap shifts inwards
and the potential drop, $\Psi(s_2)$, reduces significantly.
For example, we obtain $\Psi(s_2)=7.1 \times 10^{12}$~V for case~4,
whereas $1.7 \times 10^{13}$~V for case~2. 
A detailed physical interpretation is given in \S~\ref{sec:width}.

\subsubsection{Gamma-ray Spectra}
\label{sec:res_spec}

We compute the GeV and TeV spectrum by the method described in 
\S~\ref{sec:Gamma-ray_spc}.
We adopt the cross sectional area of
$D_\perp{}^2= (6 W)^2$ for all the cases to be considered, 
so that the GeV flux in cases~1 and 2 may be consistent with observations.
If $D_\perp$ increase twice, both the GeV and TeV fluxes 
increases four times.

\begin{figure} 
\centerline{ \epsfxsize=8.5cm \epsfbox[200 20 500 250]
   {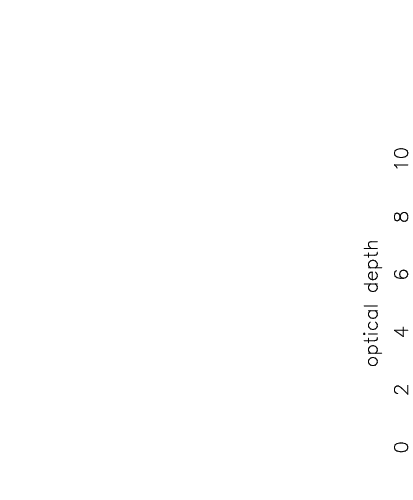} } 
\caption{\label{fig:Tau_Crab_45_SSA} 
Pair-production optical depth for the IR field represented by
SSA spectrum (case~A).
        }
\end{figure}

First, we consider case~A in which
the IR spectrum is approximated by SSA with turnover frequency
$\sim 1.2 \times 10^{14}$~Hz. 
In this case, 
the pair-production optical depth $\tau$ computed from
equation~(\ref{eq:tauTeV}) becomes
as presented in figure~\ref{fig:Tau_Crab_45_SSA}.
This result indicates that the TeV flux is significantly absorbed 
above $1$~TeV.

\begin{figure} 
\centerline{ \epsfxsize=8.5cm \epsfbox[200 20 500 250]
   {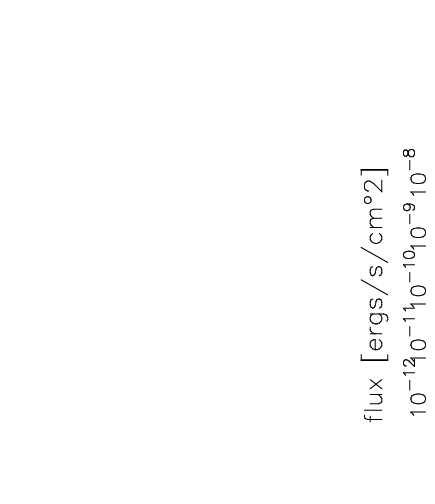} } 
\caption{\label{fig:Sp1_Crab_45_SSA} 
Spectra of the outwardly propagating $\gamma$-rays emitted from 
the Crab pulsar magnetosphere
when the IR field is approximated by a SSA spectrum (case~A).
The solid, dashed, dash-dotted, and dotted curves correspond to 
the same boundary conditions as in figure~2.
        }
\end{figure}

\begin{figure} 
\centerline{ \epsfxsize=8.5cm \epsfbox[200 20 500 250]
   {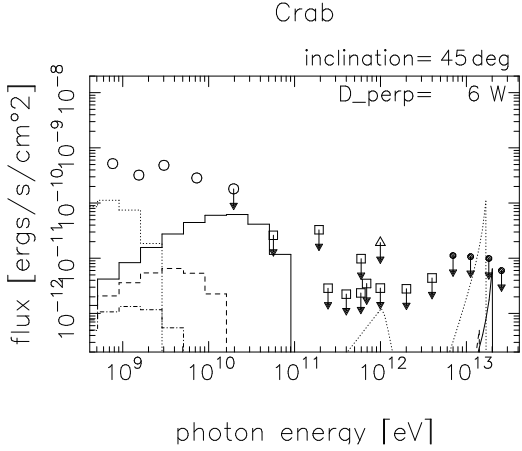} } 
\caption{\label{fig:Sp2_Crab_45_SSA} 
Same figure as figure~4 but the $\gamma$-rays are inwardly propagating. 
        }
\end{figure}

For the four different boundary conditions (cases~1, 2, 3, and 4), 
we present the spectra of the outwardly and inwardly propagating
$\gamma$-rays in figures~\ref{fig:Sp1_Crab_45_SSA} and
\ref{fig:Sp2_Crab_45_SSA}, respectively.
In GeV energies,
the observational pulsed spectrum is obtained by
EGRET observations (open circles; Nolan et al. 1993),
while in TeV energies,
only the upper limits are obtained 
by Whipple observations 
(open squares; Weekes et al. 1989; Reynolds et al. 1993;
 Goret et al. 1993; Hillas, A. M.; Lessard et al. 2000),
Durham observations (open triangle; Dowthwaite et al. 1984),
and CELESTE observations 
(open square at 60~GeV; Holder, J., private communication).
The filled circles denote the unpulsed flux obtained by 
CANGAROO observations (Tanimori et al. 1998).

It follows from figures~\ref{fig:Sp1_Crab_45_SSA} 
and \ref{fig:Sp2_Crab_45_SSA} that
the TeV flux is undetectable except for $h\nu \sim 10$~TeV.
Around $10$~TeV,
the $\gamma$-ray flux is slightly less than or comparable with 
the observational upper limits for
cases~1, 2, and 3,
and exceeds the limits for case~4.
Nevertheless, we can exclude case~4 from consideration,
because the expected GeV spectrum is very very soft and
is inconsistent with the EGRET observations,
whatever $D_\perp$ we may assume.

It is noteworthy that the GeV spectrum,
which does not depend on the assumed IR field, 
depends on $j_1$ and $j_2$ significantly.
In particular, in case~4 (as the dotted curves show), 
the GeV emission significantly decreases and softens,
because both the potential drop and the maximum of $\Ell$ reduce
as the gap shifts inwards.
As a result, it becomes impossible to explain the EGRET flux
around $10$~GeV,
if the gap is located well inside of the conventional null surface.

\begin{figure} 
\centerline{ \epsfxsize=8.5cm \epsfbox[200 20 500 250]
   {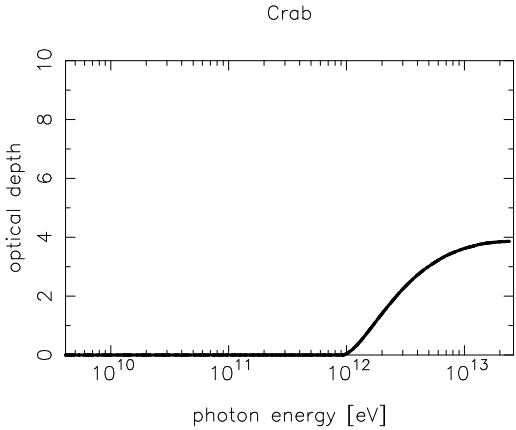} } 
\caption{\label{fig:Tau_Crab_45_int1} 
Pair-production optical depth 
for the single power-law IR spectrum (fig.~1) fitted
between radio and optical bands (case~B).
        }
\end{figure}

Next, let us next consider the case~B in which 
the IR spectrum is interpolated from radio and optical pulsed fluxes.
In this case, 
the pair-production optical depth $\tau$ computed from
equation(~\ref{eq:tauTeV}) becomes
as presented in figure~\ref{fig:Tau_Crab_45_int1}.
Therefore, the emitted TeV flux
significantly reduces above $1$~TeV.

\begin{figure} 
\centerline{ \epsfxsize=8.5cm \epsfbox[200 20 500 250]
   {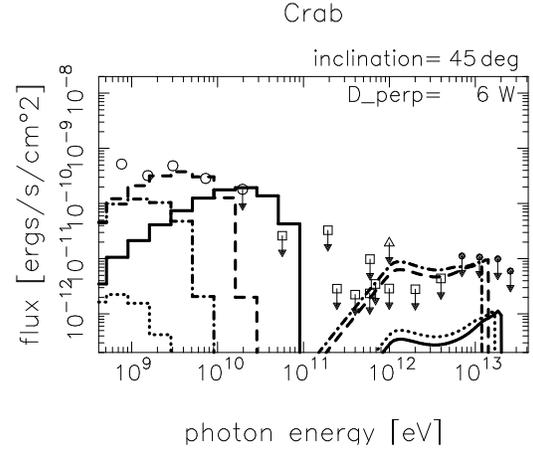} } 
\caption{\label{fig:Sp1_Crab_45_int1} 
Spectra of outwardly propagating $\gamma$-rays
from the Crab pulsar magnetosphere
when the IR spectrum is interpolated from radio and optical bands 
with a single power law (case~B).
The solid, dashed, dash-dotted, and dotted curves correspond to 
the same cases as in figure~2.
        }
\end{figure}

\begin{figure} 
\centerline{ \epsfxsize=8.5cm \epsfbox[200 20 500 250]
   {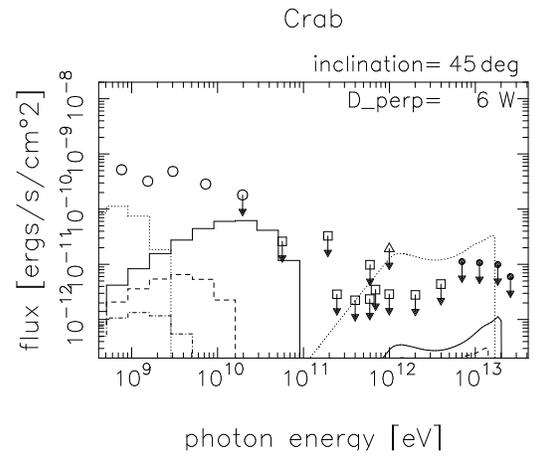} } 
\caption{\label{fig:Sp2_Crab_45_int1} 
Same figure as figure~6
but the $\gamma$-rays are inwardly propagating.
        }
\end{figure}

The spectra of the outwardly and inwardly propagating $\gamma$-rays are
presented in figures~\ref{fig:Sp1_Crab_45_int1} and
\ref{fig:Sp2_Crab_45_int1}, respectively.
It follows from the two figures that
the TeV fluxes exceed the observational upper limits for
cases 2, 3, and 4.
In case~1, the upscattered flux is small because of
its small $N_{\rm e}$, which is proportional to $j_{\rm tot}= 0.01$.

\begin{figure} 
\centerline{ \epsfxsize=8.5cm \epsfbox[200 20 500 250]
   {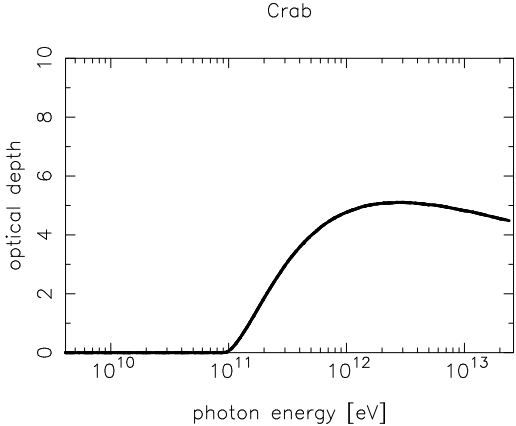} } 
\caption{\label{fig:Tau_Crab_45_int2} 
Same figure as fig.~6 
but with a large upper cutoff energy,
$\epsilon_{\rm IR,max}=10^{-5}$.
        }
\end{figure}

\begin{figure} 
\centerline{ \epsfxsize=8.5cm \epsfbox[200 20 500 250]
   {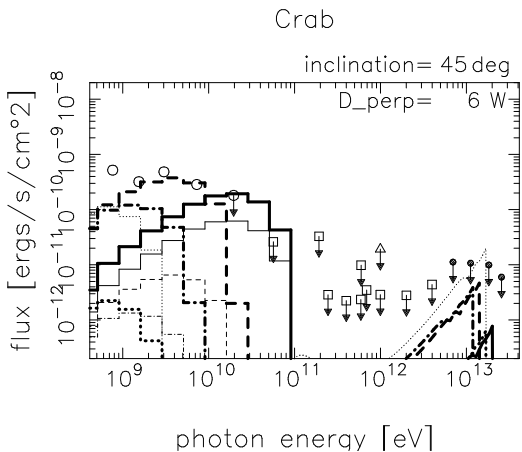} } 
\caption{\label{fig:Sp3_Crab_45_int2} 
Gamma-ray spectra from the Crab pulsar magnetosphere
for case~B but with a large upper cutoff energy,
$\epsilon_{\rm IR,max}=10^{-5}$.
The thick (or thin) curves denote outwardly (or inwardly) propagating
$\gamma$-rays.
        }
\end{figure}

Let us briefly consider the case when the interpolated spectrum
($N_0= 1.5 \times 10^{17}$ and $\alpha=-0.88$)
extends to much higher frequencies
and adopt $\epsilon_{\rm IR,max}= 10^{-5}$ (or $1.2 \times 10^{15}$~Hz).
In this case, $\tau>2$ holds above $0.2$~TeV 
(fig.~\ref{fig:Tau_Crab_45_int2});
the absorbed TeV flux is thus suppressed below the observational 
upper limits as figure~\ref{fig:Sp3_Crab_45_int2} indicates.

In short, we can conclude that the problem of the excessive TeV flux
does not arise 
if the IR field is represented by a SSA spectrum (case~A)
or if the IR field is interpolated by a single power law (case~B)
with a large cut-off energy ($\epsilon_{\rm IR,max} \sim 10^{-5}$).

\subsection{Dependence on Magnetic Inclination}
\label{sec:inclination}

\begin{figure} 
\centerline{ \epsfxsize=8.5cm \epsfbox[200 20 500 250]
   {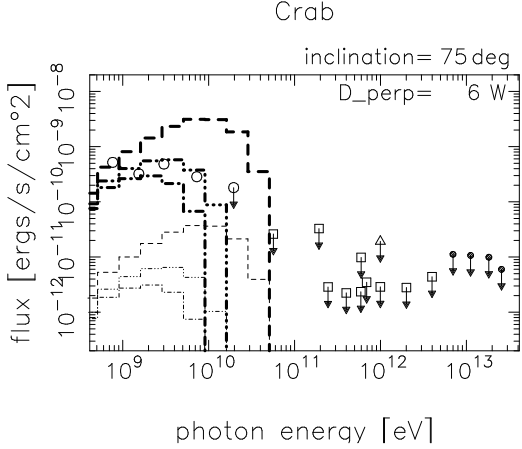} } 
\caption{\label{fig:Sp3_Crab_75_SSA} 
Gamma-ray spectra from the Crab pulsar magnetosphere
with $\inc=75^\circ$, 
when the IR field is approximated by a SSA spectrum (case~A).
The thick (or thin) curves denote outwardly (or inwardly) propagating
$\gamma$-rays.
        }
\end{figure}

To investigate how the results depend on the magnetic inclination,
we present the expected Crab pulsar spectra for 
$\inc=75^\circ$ in figure~\ref{fig:Sp3_Crab_75_SSA}.
The dashed and dash-dotted curves correspond to cases~2 and 3,
while the dash-dot-dot-dotted ones to the case of
$j_1=0.5$ and $j_2=0$.
Case~1 is not depicted because the central energy 
of curvature-radiated
photons becomes comparable with $\beta_{11} m_{\rm e}c^2= 90.8$~GeV;
in this case, its hard spectrum would be inconsistent with the
EGRET pulsed spectrum below 30~GeV and the CELESTE upper limit at 60~GeV.
Moreover, case~4 (i.e., $j_1=0$ and $j_2=0.3$) is excluded 
in figure~\ref{fig:Sp3_Crab_75_SSA};
this is because the gap is located so close to the star surface that
the IC scatterings dominates the curvature process. 

Comparing the GeV spectra in figure~\ref{fig:Sp3_Crab_75_SSA}
with those in 
figures~\ref{fig:Sp1_Crab_45_SSA} and \ref{fig:Sp2_Crab_45_SSA},
we can confirm that the curvature emission becomes hard and 
luminous if $\inc$ increases.
Its physical interpretation will be discussed in \S~\ref{sec:interpret}.
In figure~\ref{fig:Sp3_Crab_75_SSA}, 
case~A is adopted as the infrared spectrum;
however, the IR field is not important 
when we discuss the curvature-radiated $\gamma$-ray spectrum. 

It also follows from figure~\ref{fig:Sp3_Crab_75_SSA}
that the observed, pulsed GeV spectrum can be explained
if we take $j_1=0.5$ and $j_2=0$ for $\inc=75^\circ$.
In other words, the curvature spectrum becomes analogous
between $j_1=0.3$ for $\inc=45^\circ$ 
(dashed curve in fig.~\ref{fig:Sp1_Crab_45_SSA}) 
and $j_1=0.5$ for $\inc=75^\circ$
(thick, dash-dot-dot-dotted one in fig.~\ref{fig:Sp3_Crab_75_SSA}), 
if we fix $j_{\rm gap}=0.01$ and $j_2=0$.
That is, a greater $j_1$ is preferable for a greater $\inc$.
It is natural, because the decrease of the distance
of the intersection between the conventional null surface and the
last-open field line from the star surface 
(with increasing $\inc$)
should be compensated by shifting the gap outwards
(with increasing $j_1$),
so that the gap may have the comparable magnetic and X-ray field 
strengths.

On these grouds, we can conclude that we cannot decouple 
the effects of the magnetospheric currents ($j_1$,$j_2$) and $\inc$,
if we only compare the fluxes of the outwardly propagating $\gamma$-rays
(when $j_1>j_2$).
It would be possible to argue that these two effects could be
decoupled if we considered the inward/outward flux ratio,
or the three-dimensional structure of the accelerator.
However, such details are irrelevant to the main subject of this paper.

\section{Discussion}
\label{sec:discussion}

In summary, we have developed a one-dimensional model for 
an outer-gap accelerator in the magnetosphere of a rotation-powered pulsar.
When a magnetospheric current flows into the gap
from the outer (or inner) boundary,
the gap shifts inwards (or outwards).
In particular, 
when a good fraction of the Goldreich-Julian current density is
injected from the outer boundary,
the gap is located well inside of the conventional null surface;
the resultant GeV emission becomes very soft and weak.
Applying this method to the Crab pulsar, 
we find that the gap should be located near to or outside of the 
conventional null surface,
so that the observed GeV spectrum of pulsed GeV fluxes may be emitted
via a curvature process.
By virtue of the absorption by the dense IR field in the magnetosphere, 
the problem of excessive TeV emission does not arise.

\subsection{Gap Width vs. Current Injection}
\label{sec:width}

By utilizing the gap closure condition~(\ref{eq:closure}),
we can interpret why $W$ becomes significantly small
when the gap is located well inside of the conventional null surface
(\S~\ref{sec:res_Ell}). 
First, the X-ray density becomes large at small radii
to reduce $\lambda_{\rm p}$ in equation~(\ref{eq:closure}).
Secondly, the ratio $j_{\rm gap}$/$j_{\rm tot}$ decreases
as $j_2$ increases.
As a result, $W$ decreases very rapidly with increasing $j_2$.
When $W$ decreases, $N_\gamma$ decreases to some extent;
however, this effect is passive and cannot change the conclusion.
On these grounds, the gap width significantly decreases
when particles are injected at the {\it outer} boundary.
Therefore, the potential drop also decreases significantly.
 
On the other hand, when the gap is located outside of the 
conventional null surface,
the decreased $j_{\rm gap}/j_{\rm tot}$ due to the increase of $j_1$
partially cancels with the increase of $\lambda_{\rm p}$ due to
the diluted X-ray field.
Thus, the gap width is roughly unchanged when particles are
injected at the {\it inner} boundary.

\subsection{Interpretation of the Magnetic-Inclination Dependence}
\label{sec:interpret}

In this subsection, we interpret the dependence of the results
on $\inc$ (in \S~\ref{sec:inclination}).
In Pape~V, it was predicted that $W$ ($=2H$ in their notation)
is a decreasing function of $\inc$ for all the twelve pulsars
considered.
The reasons are fivefolds:\\
$\bullet$ With no current injection 
(i.e., $j_1=j_2=0$ as considered in Paper~V),
the gap is located at the intersection of the last-open field line 
and the conventional null surface, where $B_z$ vanishes.\\
$\bullet$ The intersection approaches the star if $\inc$ increases.\\
$\bullet$ The density of the X-ray field illuminating the gap increases
as the intersection approaches the star
(or equivalently, as $r_{\rm cnt}$ decreases).\\
$\bullet$ It follows from the closure condition (eq.~[\ref{eq:closure}]) 
that $W \propto \lambda_{\rm p}/N_\gamma$.
If we neglect the variations in $N_\gamma$, 
$W$ is proportional to
$\lambda_{\rm p} \propto N_{\rm x}{}^{-1} \propto r_{\rm cnt}{}^{-2}$,
where $N_{\rm x}$ is the X-ray number density.
Therefore, $W$ decreases with decreasing $r_{\rm cnt}$, 
and hence with increasing $\inc$.\\
$\bullet$ In reality, $N_\gamma$ decreases if $W$ decreases.
As a result of this \lq negative feedback effect',
the decrease of $W$ for an increasing $\inc$ is partially canceled.
However, this effect is passive; therefore, the conclusion of the
decreasing $W$ with increasing $\inc$ is unchanged.

In Paper~V,
the GeV emission is predicted to become hard and luminous,
as $\inc$ increases for the same set of $j_{\rm gap}$, $j_1$, and $j_2$.
The reasons are fivefold:\\
$\bullet$ The gap approaches the star
(i.e., $r_{\rm cnt}$ decreases), 
as $\inc$ increases for fixed $j_1$ and $j_2$
(say, $j_1=j_2=0$ as considered in Paper~V).\\
$\bullet$ The magnetic field in the gap increases as 
$B \propto r_{\rm cnt}{}^{-3}$ 
when the gap approaches the star.\\
$\bullet$ As a result of this rapid increase of $B$,
$\Ell$ increases (e.g., eq.~[\ref{eq:Poisson_0}]),
in spite of the decreasing $W$, as stated in the paragraph just above.\\
$\bullet$ The increased $\Ell$ for a larger $\inc$ results in a
harder curvature spectrum in GeV energies.\\
$\bullet$ The potential drop in the gap 
is roughly proportional to the maximum of $\Ell$ in the gap
times $W$.
Because of the \lq negative feedback effect' due to $N_\gamma$,
the weakly decreasing $W$ cannot cancel the increase of $\Ell$.
As a result, the potential drop, and hence the GeV luminosity 
increases with increasing $\inc$.

\subsection{Super Goldreich-Julian Current}
\label{sec:solution_space}

\begin{figure} 
\centerline{ \epsfxsize=8.5cm \epsfbox[200 20 500 250]
             {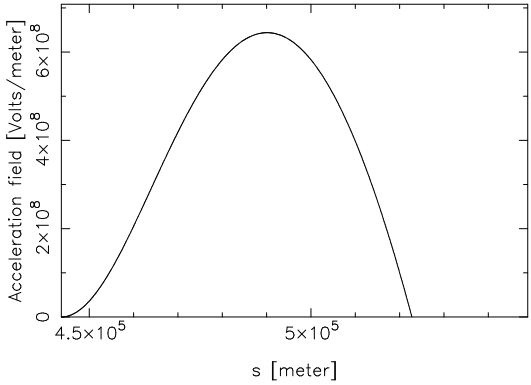} } 
\caption{\label{fig:Ell_Crab_45_Jmax} 
Distribution of $\Ell(s)$ for the Crab pulsar with $\inc=45^\circ$
when $j_{\rm gap}=0.0255$, $j_1=j_2=0$. 
        }
\end{figure}

In this subsection, we briefly discuss the relaxation of the
limit of the current density flowing in the gap along the field lines.
In Papers~I, II, III, VI, in which $j_1=j_2=0$ is assumed,
stationary gap solutions were found
only for a small $j_{\rm gap}$.
By the revised method presented in this paper,
the solutions for the Crab pulsar exist for 
$j_{\rm gap} < 0.0255$, if we set $j_1=j_2=0$. 
The solution of $\Ell(s)$ for $j_1=j_2=0$ and $j_{\rm gap}=0.0255$
for the Crab pulsar when $\inc=45^\circ$ is depicted 
in figure~\ref{fig:Ell_Crab_45_Jmax}.
Because of the \lq brim' at the inner boundary,
no solution exists for $j_{\rm gap}>0.0255$.
In this case, $j_{\rm tot}= j_{\rm gap}+j_1+j_2$ 
is limited only below $0.0255$,
which is much less than the typical Goldreich-Julian value, $1$.

Let us briefly consider how much $j_{\rm tot}$ is needed 
for the observed spin-down luminosity to be emitted.
If we assume that all the current flowing in the magnetosphere
penetrate the gap, then the net current becomes
$J= (\Omega/2\pi)j_{\rm tot}\Psi$, 
where $\Psi$ is the magnetic fluxes along which the current is flowing.
Assuming a magnetic dipole radiation, 
we obtain the potential drop at the stellar surface as
$V_\ast \sim \Omega \Psi/(\pi c)$.
The spin-down luminosity then becomes
\begin{equation}
  \dot{E}_{\rm rot} 
  = J \times V_\ast
  \sim \frac{j_{\rm tot}}{c}
       \left( \frac{\Omega}{\pi}\Psi \right)^2.
  \label{eq:spin_down}
\end{equation}
If the gap is geometrically thick in the transfield directions,
we may expect that the field lines thread the polar cap 
with area $A_{\rm pole} \equiv \pi (r_\ast \sin\theta_\ast)^2$, 
where $r_\ast$ refers to the stellar radius and
$\theta_\ast$ to the colatitude angle between the magnetic axis 
and the last-open field line.
Utilizing $\sin^2\theta_\ast/r_\ast = \mbox{constant} \sim \Omega/c$
for a dipole geometry, we obtain
\begin{equation}
  \Psi \sim \frac{\mu_{\rm m}}{r_\ast^3} A_{\rm pole}
       \sim \pi \frac{\Omega\mu_{\rm m}}{c},
  \label{eq:Psi}
\end{equation}
where $\mu_{\rm m}$ is the neutron star's magnetic dipole moment.
Substituting equation~(\ref{eq:Psi}) into (\ref{eq:spin_down}),
we obtain
\begin{equation}
  \dot{E}_{\rm rot} 
  \sim j_{\rm tot} \frac{\Omega^4 \mu_{\rm m}^2}{c^3}
\end{equation}
For the Crab pulsar, 
$\Omega^4 \mu_{\rm m}^2 / c^3 = 10^{38.6} \mbox{ergs s}^{-1}$; 
therefore, $j_{\rm tot} \sim 1$ is required, 
so that the observed spin-down luminosity 
$10^{38.65} \mbox{ergs s}^{-1}$ may be realized.
Analogous conclusions are derived for other rotation-powered pulsars.
Moreover, the sharp pulse of the Crab pulsar may imply 
$A_{\rm pole} \ll \pi (r_\ast \sin\theta_\ast)^2$;
therefore, even $j_{\rm gap} \gg 1$ may be required.
On these grounds, the limitation of 
$j_{\rm tot}= j_{\rm gap} \ll 1$ 
derived for $j_1= j_2=0$ were insufficient to apply to realistic pulsars.

In the present paper, we relaxed the limitation of $j_{\rm tot}$
by allowing $j_1$ or $j_2$ to be non-vanishing.
The results of the predicted $\gamma$-ray spectra are, therefore, 
more realistic compared with previous results 
obtained in Papers~I, II, III, VI.
However, even in this treatment, 
$j_{\rm tot}$ is limited below unity.

The next issue is, therefore, 
to consider whether we can construct an outer-gap model
with super-Goldreich-Julian current density (i.e., $j_{\rm gap}>1$).
The Poisson equation~(\ref{eq:Poisson_1}) tells that solutions
exit even for $j_1+j_2 \gg 1$, provided that $j_1-j_2 <1$.
(For example, if $j_1=j_2 \gg 1$, the gap exists at the
 conventional null surface.)
In this case, $W$ becomes much smaller than those obtained for
$j_1+j_2<1$  because of the gap closure condition 
(eq.~[\ref{eq:closure}]). 
In the case of the Crab pulsar, 
the small $W$ obtained for $j_1+j_2>1$
fails the mono-enegetic approximation.
To find solutions for $j_{\rm tot} \sim j_1+j_2 \gg 1$,
we could assume much smaller collision angles 
so that the pair-production mean free path may become much larger.
To settle this issue, we must constrain the magnetic field geometry 
around the gap and quantitatively infer the collision angles between 
the primary $\gamma$-rays and the secondary X-rays.

In short, stationary gap solutions exit 
even for a super Goldreich-Julian current.
In this case, the collision angles should be much less than
$W/\rlc$ so that the emitted $\gamma$-ray flux may be 
consistent with observations 
for the young pulsars whose X-ray field is dense (like Crab).
For older pulsars whose X-ray field is less dense, on the other hand,
we can in fact find solutions with super Goldreich-Julian current.
It will be discussed in a subsequent paper.

\subsection{Comparison with Previous Works}
\label{sec:comparison}

Let us compare the present methods and results with Paper~V. 
In the present paper, $\Ell$, $N_\pm(s)$, and $G_\pm(s,\epsilon_\gamma)$ 
were solved from the Vlasov equations for a non-vacuum gap, 
while in Paper~V only $\Ell$ field was solved from the Poisson equation
for a vacuum gap,
with the aid of $W$, which was deduced from the gap closure condition.
In the stationary gap, the Vlasov equations automatically satisfy the
closure condition; 
therefore, the obtained electrodynamic structures 
(e.g., $W$, $\Ell$) are essentially the same between the two Papers, 
provided that the gap is nearly vacuum (i.e., $j_{\rm tot} \ll 1$).
By relaxing the boundary conditions of the magnetospheric current,
and by solving the non-vacuum solution from the Vlasov equations,
we first find in this paper 
an interesting behavior of the gap position:
The gap shifts outwards (or inwards) when current is injected
from the inner (or outer) boundary.
The obtained GeV spectra are similar between the two papers,
unless the gap is located well inside of the conventional null surface. 
In Paper~V, the intrinsic TeV spectra were depicted in figure~6;
on the other hand, in this paper, the TeV spectra after absorption
were depicted in figures~\ref{fig:Sp1_Crab_45_SSA}, 
\ref{fig:Sp2_Crab_45_SSA},
\ref{fig:Sp1_Crab_45_int1},
\ref{fig:Sp2_Crab_45_int1}, and 
\ref{fig:Sp3_Crab_45_int2}.

We briefly compare the present method with ZC97,
who considered that the gap width is limited by the
surface X-rays due to the bombardment of the particles produced in the gap.
The magnetospheric X-rays considered in this paper
is much denser than the surface X-rays due to the bombardment. 
As a result, the localized gap in the present paper produces less
intrinsic TeV flux compared with what would be obtained in ZC97 picture.

\subsection{Possibility of Another Solution Branch}
\label{sec:another_branch}

For cases~1, 2, and 3, 
the intrinsic TeV luminosity is comparable or less than the GeV one.
Therefore, the Lorentz factors are limited primarily by the curvature 
process (eq.[\ref{eq:terminal}]).
For case~4, however,
the intrinsic TeV luminosity well exceeds the GeV one;
therefore, the radiation-reaction forces are due to IC scatterings
rather than the curvature process.
In fact, we may expect a sufficient GeV flux via IC scatterings 
when the gap is located well inside of the conventional null surface.
This is because the dense X-ray field will limit the particle Lorentz 
factors small (Paper~II),
and because the less-energetic particles scatter 
copious IR photons into lower $\gamma$-ray energies
with large cross sections ($\sim \sgT$).
There is room for further investigation on this issue.

\par
\vspace{1pc}\par

One of the authors (K. H.) wishes to express his gratitude to
Drs. Y. Saito and A. K. Harding for valuable advice. 
He also thanks the Astronomical Data Analysis Center of
National Astronomical Observatory, Japan for the use of workstations.

\end{document}